\documentclass[12pt]{article}
\usepackage{amsthm,amssymb,amsmath,url,amsfonts,mathrsfs}
\newcounter{myfn}
\title{On the Approach to Thermal Equilibrium of Macroscopic Quantum Systems}
\author{
Sheldon Goldstein,\footnote{Departments of Mathematics and Physics, 
	Rutgers University, 
	110 Frelinghuysen Road, Piscataway, NJ 08854-8019, USA.}
	{}\footnote{E-mail: oldstein@math.rutgers.edu}\ \ 
Joel L. Lebowitz,\setcounter{myfn}{1}\textsuperscript{\fnsymbol{myfn}}{}\footnote{E-mail: 
	lebowitz@math.rutgers.edu}\ \ 
Christian Mastrodonato,\footnote{Dipartimento di Fisica dell'Universit\`a di
	Genova and INFN sezione di Genova, Via Dodecaneso 33, 16146
	Genova, Italy.}\ \footnote{E-mail: christian.mastrodonato@ge.infn.it}\\
Roderich Tumulka,\footnote{Department of Mathematics, Rutgers University, 
	110 Frelinghuysen Road, Piscataway, NJ 08854-8019, USA.}{}\ \footnote{E-mail: 
	tumulka@math.rutgers.edu}{}\ \ \ and
Nino Zangh\`\i\setcounter{myfn}{4}\textsuperscript{\fnsymbol{myfn}}{}\footnote{E-mail: 
	zanghi@ge.infn.it}
}
\date{November 7, 2009}

\theoremstyle{plain}
\newtheorem{thm}{Theorem}
\newtheorem{lem}{Lemma}

\newcommand{\ket}[1]{\vert#1\rangle}
\newcommand{\bra}[1]{\langle#1\vert}
\newcommand{\Tr}{\text{Tr}}
\newcommand{\CCC}{\mathbb{C}}
\newcommand{\RRR}{\mathbb{R}}
\newcommand{\EEE}{\mathbb{E}}

\newcommand{\PPP}{\mathbb{P}}
\newcommand{\scp}[2]{\langle #1| #2 \rangle}
\newcommand{\Hilbert}{\mathscr{H}}
\newcommand{\be}{\begin{equation}}
\newcommand{\ee}{\end{equation}}
\renewcommand{\Re}{\mathrm{Re}}
\renewcommand{\Im}{\mathrm{Im}}
\newcommand{\D}{D} 
\newcommand{\dd}{d} 
\newcommand{\Unitary}{\mathcal{U}} 
\newcommand{\Set}{\mathscr{S}}
\newcommand{\set}{M}
\newcommand{\vq}{\boldsymbol{q}}
\newcommand{\vp}{\boldsymbol{p}}
\newcommand{\A}{\mathscr{A}}

\begin{document}

\maketitle

\begin{abstract}
We consider an isolated, macroscopic quantum system.
Let $\Hilbert$ be a micro-canonical ``energy shell,'' i.e., a subspace of the system's Hilbert space spanned by the (finitely) many energy eigenstates with energies between $E$ and $E+\delta E$. The thermal equilibrium macro-state at energy $E$ corresponds to a subspace $\Hilbert_{eq}$ of $\Hilbert$ such that $\dim\Hilbert_{eq}/\dim\Hilbert$ is close to 1. We say that a system with state vector $\psi\in\Hilbert$ is in thermal equilibrium if $\psi$ is ``close'' to $\Hilbert_{eq}$. We show that for ``typical'' Hamiltonians with given eigenvalues, all initial state vectors $\psi_0$ evolve in such a way that $\psi_t$ is in thermal equilibrium for most times $t$.  This result is closely related to von Neumann's quantum ergodic theorem of 1929.

\medskip

\end{abstract}

\section{Introduction}

If a hot brick is brought in contact with a cold brick, and the two bricks are otherwise isolated, then energy will flow from the hot to the cold brick until their temperatures become equal, i.e., the system equilibrates. Since the bricks ultimately consist of electrons and nuclei, they form a quantum system with a huge number ($>10^{20}$) of particles; this is an example of an \emph{isolated, macroscopic quantum system}. 

From a microscopic point of view the state of the system at time $t$ is described by a vector 
\be
\psi(t)=e^{-iHt}\psi(0)
\ee
in the system's Hilbert space or a density matrix
\be
\rho(t)=e^{-iHt}\rho(0)e^{iHt}\,,
\ee
where $H$ is the Hamiltonian of the isolated system and we have set $\hbar=1$. In this paper we prove a theorem asserting that for a sufficiently large quantum system with a ``typical'' Hamiltonian and an \emph{arbitrary initial state} $\psi(0)$, the system's state $\psi(t)$ spends most of the time, in the long run, in thermal equilibrium. (Of course, before the system even reaches thermal equilibrium there could be a waiting time longer than the present age of the universe.) This implies the same behavior for an arbitrary $\rho(0)$.

This behavior of isolated, macroscopic quantum systems is an instance of a phenomenon we call \emph{normal typicality} \cite{GLMTZ09}, a version of which is expressed in von Neumann's quantum ergodic theorem \cite{vN29}. However, our result falls outside the scope of von Neumann's theorem, because of the technical assumptions made in that theorem. Our result also differs from the related results in \cite{Deu91,Sre94,T98,R08,RDO08,LPSW08}, which use different notions of when a system is in an equilibrium state. In particular they do not regard the thermal equilibrium of an isolated macroscopic system as corresponding to its wave function being close to a subspace $\Hilbert_{eq}$ of Hilbert space. See Section 6 for further discussion.

The rest of this paper is organized as follows. In the remainder of Section 1, we define more precisely what we mean by thermal equilibrium. In Section 2 we outline the problem and our result, Theorem 1. In Section 3 we prove the key estimate for the proof of Theorem 1. In Section 4 we describe examples of exceptional Hamiltonians, illustrating how a system can fail to ever approach thermal equilibrium. In Section 5 we compare our result to the situation with classical systems. In Section 6 we discuss related works.

\subsection{The equilibrium subspace}
\label{sec:subspace}

Let $\Hilbert_\mathrm{total}$ be the Hilbert space of a macroscopic system in a box $\Lambda$, and let $H$ be its Hamiltonian. Let $\{\phi_\alpha\}$ be an orthonormal basis of $\Hilbert_\mathrm{total}$ consisting of eigenvectors of $H$ with eigenvalues $E_\alpha$. Consider an energy interval $[E,E+\delta E]$, where $\delta E$ is small on the macroscopic scale but large enough for the interval $[E,E+\delta E]$ to contain very many eigenvalues. Let $\Hilbert\subseteq \Hilbert_\mathrm{total}$ be the corresponding subspace,
\be
\Hilbert = \text{span}\bigl\{\phi_\alpha: E_\alpha\in[E,E+\delta E]\bigr\}\,.
\ee
A subspace such as $\Hilbert$ is often called a \emph{micro-canonical energy shell.} Let $\D$ be the dimension of $\Hilbert$, i.e., the number of energy levels, including multiplicities, between $E$ and $E+\delta E$. In the following we consider only quantum states $\psi$ that lie in $\Hilbert$, i.e., of the form
\be
\psi=\sum_{\alpha} c_\alpha \, \phi_\alpha
\ee
with $c_\alpha\neq 0$ only for $\alpha$ such that $E_\alpha\in [E,E+\delta E]$.

According to the analysis of von Neumann \cite{vN29,vN32} and others (cf.~\cite{J63}), the macroscopic (coarse-grained) observables in a macroscopic quantum system can be naturally ``rounded'' to form a set of commuting operators,
\begin{equation}\label{eq:macroObservables}
\bigl\{M_i\bigr\}_{i=1,\ldots,k}\, .
\end{equation}
The operators are defined on $\Hilbert_\mathrm{total}$, but since we can take them to include (and thus commute with) a coarse-grained Hamiltonian, we can (and will) take them to commute with the projection to $\Hilbert$, and thus to map $\Hilbert$ to itself. We write $\nu=(m_1,\ldots,m_k)$ for a list of eigenvalues $m_i$ of the restriction of $M_i$ to $\Hilbert$, and $\Hilbert_\nu$ for the joint eigenspace. Such a set of operators generates an orthogonal decomposition of the Hilbert space
\begin{equation}\label{eq:orthoDecom}
\Hilbert = \bigoplus_\nu \Hilbert_\nu  \, ,
\end{equation}
where each $\Hilbert_\nu$, called a macro-space, represents a macro-state of the
system. The dimension of $\Hilbert_\nu$ is denoted by $\dd_\nu$; note that $\sum_\nu \dd_\nu = \D$.  If any $\Hilbert_\nu$ has dimension 0, we remove it from the family $\{\Hilbert_\nu\}$. In practice, $\dd_\nu \gg 1$, since we are considering a macroscopic system with coarse-grained observables.

It can be shown in many cases, and is expected to be true generally,
that among the macro-spaces $\Hilbert_\nu$ there is a particular macro-space $\Hilbert_{eq}$, the one corresponding to thermal equilibrium, such that
\be\label{deqD1}
\dd_{eq}/\D \approx 1\,,
\ee
indeed with the difference $1-\dd_{eq}/\D$ exponentially small in the number of particles.\footnote{This dominance of the equilibrium state can be expressed in terms of the (Boltzmann) entropy $S_\nu$ of a macroscopic system in the macro-state $\nu$, be it the equilibrium state or some other (see \cite{L07}), defined as $S_\nu = k_B \log \dd_\nu$, where $k_B$ is the Boltzmann constant: $\dd_{eq}/\D$ being close to
1 just expresses the fact that the entropy of the equilibrium state is close to the micro-canonical entropy $S_{mc}$, i.e., $S_{eq} = k_B \log \dd_{eq} \approx k_B \log \D = S_{mc}$.} 
This implies, in particular, that each of the macro-observables $M_i$ is ``nearly constant'' on the energy shell $\Hilbert$ in the sense that one of its eigenvalues has multiplicity at least $\dd_{eq}\approx \D$. 
We say that a system with quantum state $\psi$ (with $\|\psi\|=1$) is \emph{in thermal equilibrium}  if $\psi$ is very close (in the Hilbert space norm) to $\Hilbert_{eq}$, i.e., if
\be\label{eq1}
\scp{\psi}{P_{eq}|\psi} \approx 1\,, 
\ee
where $P_{eq}$ is the projection operator to $\Hilbert_{eq}$. The corresponding relation for density matrices is
\be
\Tr(P_{eq} \rho) \approx 1\,.
\ee
The condition \eqref{eq1} implies that a quantum measurement of the macroscopic observable $M_i$ on a system with wave function $\psi$ will yield, with probability close to 1, the ``equilibrium'' value of $M_i$. Likewise, a joint measurement of $M_1,\ldots,M_k$ will yield, with probability close to 1, their equilibrium values.

Let $\mu(d\psi)$ be the uniform measure on the unit sphere in $\Hilbert$ \cite{schrbook,Bloch}. It follows from \eqref{deqD1} that most $\psi$ relative to $\mu$ are in thermal equilibrium. Indeed,
\begin{equation}\label{eq:mostVectors}
\int \bra{\psi}P_{eq}\ket{\psi}\, \mu(d\psi) = \frac{1}{\D}\Tr\,P_{eq} 
= \frac{\dd_{eq}}{\D} \approx 1 \,.
\end{equation}
Since the quantity $\bra{\psi}P_{eq}\ket{\psi}$ is bounded from above by 1, most $\psi$ must satisfy \eqref{eq1}.\footnote{It should in fact be true for a large class of observables $A$ on $\Hilbert$ that, for most $\psi$ relative to $\mu$, $\scp{\psi}{A|\psi} \approx \Tr(\rho_{mc}A)$, where $\rho_{mc}$ is the micro-canonical density matrix, i.e., $1/\D$ times the identity on $\Hilbert$. This is relevant to the various results on thermalization described in Section~\ref{sec:history}.}

\subsection{Examples of equilibrium subspaces}\label{sec:example}

To illustrate the decomposition into macro-states, we describe two examples. As Example 1, consider a system composed of two identical subsystems designated 1 and 2, e.g., the bricks mentioned in the beginning of this paper, with Hilbert space $\Hilbert_\mathrm{total} = \Hilbert_1 \otimes \Hilbert_2$. The Hamiltonian of the total system is
\begin{equation}\label{eq:hamiltonian}
H = H_1 + H_2 + \lambda V\, ,
\end{equation}
where $H_1$ and $H_2$ are the Hamiltonians of subsystems 1 and 2
respectively, and $\lambda V$ is a small interaction between the two subsystems. We assume that $H_1$, $H_2$, and $H$ are positive operators. Let $\Hilbert$ be spanned by the eigenfunctions of $H$ with energies between $E$ and $E+\delta E$. 

In this example, we consider just a single macro-observable $M$, which is a projected and coarse-grained version of $H_1/E$, i.e., of the fraction of the energy that is contained in subsystem 1 alone. We cannot take $M$ to simply equal $H_1/E$ because $H_1$ is defined on $\Hilbert_\mathrm{total}$, not $\Hilbert$, and will generically not map $\Hilbert$ to itself, while we would like $M$ to be an operator on $\Hilbert$. To obtain an operator on $\Hilbert$, let $P$ be the projection $\Hilbert_\mathrm{total}\to\Hilbert$ and set
\be
H_1'=PH_1P
\ee
(more precisely, $H_1'$ is $PH_1$ restricted to $\Hilbert$). Note that $H_1'$ is a positive operator, but might have eigenvalues greater than $E$. Now define\footnote{Recall that the application of a function $f$ to a self-adjoint matrix $A$ is defined to be $f(A)=\sum f(a_\alpha)|\varphi_\alpha\rangle\langle\varphi_\alpha|$ if the spectral decomposition of $A$ reads $A=\sum a_\alpha |\varphi_\alpha\rangle \langle\varphi_\alpha|$.}
\be
M=f(H_1'/E)
\ee
with the coarse-graining function
\be\label{fdef}
f(x)=\begin{cases}
0&\text{if }x<0.01,\\
0.02&\text{if }x\in[0.01,0.03),\\
0.04&\text{if }x\in[0.03,0.05),\\\text{etc.}&\ldots 
\end{cases}
\ee
The $\Hilbert_\nu$ are the eigenspaces of $M$; clearly, $\oplus_\nu\Hilbert_\nu = \Hilbert$. If, as we assume, $\lambda V$ is small, then we expect $\Hilbert_{0.5}=\Hilbert_{eq}$ to have the overwhelming majority of dimensions. In a thorough treatment we would need to prove this claim, as well as that $H_1'$ is not too different from $H_1$, but we do not give such a treatment here.

\bigskip

As Example 2, consider $N$ bosons (fermions) in a box $\Lambda=[0,L]^3 \subseteq \RRR^3$; i.e., $\Hilbert_\mathrm{total}$ consists of the square-integrable (anti-)symmetric functions on $\Lambda^N$. Let the Hamiltonian be
\be\label{Hex2}
H=-\frac{1}{2m} \sum_{i=1}^N \nabla_i^2 + 
\sum_{i<j} v\bigl(|\vq_i-\vq_j|\bigr)\,,
\ee
where the Laplacian $\nabla_i^2$ has Dirichlet boundary conditions, $v(r)$ is a given pair potential, and $\vq_i$ is the triple of position coordinates of the $i$-th particle. Let $\Hilbert$ again be spanned by the eigenfunctions with energies between $E$ and $E+\delta E$. 

In this example, we consider again a single macro-observable $M$, based on the operator $N_\mathrm{left}$ for the number of particles in the left half of the box $\Lambda$:
\be
N_\mathrm{left}\psi(\vq_1,\ldots,\vq_N) = 
\#\bigl\{i: \vq_i\in [0,L/2]\times [0,L]^2\bigr\} \, \psi(\vq_1,\ldots,\vq_N)\,.
\ee
Note that the spectrum of $N_\mathrm{left}$ consists of the $N+1$ eigenvalues $0,1,2,\ldots,N$. To obtain an operator on $\Hilbert$, let $P$ be the projection $\Hilbert_\mathrm{total}\to\Hilbert$ and set $N'_\mathrm{left}=PN_\mathrm{left}P$. Note that the spectrum of $N'_\mathrm{left}$ is still contained in $[0,N]$. Now define $M=f(N'_\mathrm{left}/N)$ with the coarse-graining function \eqref{fdef}. We expect that for large $N$, the eigenspace with eigenvalue $0.5$, $\Hilbert_{eq}=\Hilbert_{0.5}$, has the overwhelming majority of dimensions (and that $N'_\mathrm{left}\approx N_\mathrm{left}$).

\section{Formulation of problem and results}

Our goal is to show that, for typical macroscopic quantum systems,
\be\label{eqmostt}
\scp{\psi(t)}{P_{eq}|\psi(t)}\approx 1 \quad \text{for most }t\,.
\ee
To see this, we compute the time average of $\scp{\psi(t)}{P_{eq}|\psi(t)}$. We denote the time average of a time-dependent quantity $f(t)$ by a bar,
\be
\overline{f(t)} = \lim_{T\to\infty} \frac{1}{T} \int_0^T dt \, f(t)\,.
\ee
Since $\scp{\psi(t)}{P_{eq}|\psi(t)}$ is always a real number between 0 and 1, it follows that if its time average is close to 1 then it must be close to 1 most of the time. 
Moreover, for $\mu$-most $\psi(0)$, where $\mu$ is the uniform measure on
the unit sphere of $\Hilbert$, $\psi(t)$ is in thermal equilibrium most of
the time. This result follows from Fubini's theorem (which implies that
taking the $\mu$-average commutes with taking the time average) and the
unitary invariance of $\mu$:
\be
\int \overline{\scp{\psi(t)}{P_{eq}|\psi(t)}} \, \mu(d\psi)
= \overline{\int \scp{\psi}{e^{iHt}P_{eq}e^{-iHt}|\psi} \, \mu(d\psi)}
=\int \scp{\psi}{P_{eq}|\psi}\,\mu(d\psi) \approx 1\,.
\ee
That is, the ensemble average of the time average is near 1, so, for
$\mu$-most $\psi(0)$, the time average must be near 1, which implies our
claim above. So the interesting question is about the behavior of exceptional $\psi(0)$, e.g., of systems which are not in thermal equilibrium at $t=0$. Do they ever go to thermal equilibrium?

As we will show, for many Hamiltonians the statement \eqref{eqmostt} holds in fact for \emph{all} $\psi(0)\in\Hilbert$. From now on, let $H$ denote the restriction of the Hamiltonian to $\Hilbert$, and let $\phi_1,\ldots,\phi_\D$ be an orthonormal basis of $\Hilbert$ consisting of eigenvectors of the Hamiltonian $H$ with eigenvalues $E_1,\ldots,E_\D$. If
\be
\psi(0) = \sum_{\alpha=1}^\D c_\alpha \, \phi_\alpha\,,
\quad c_\alpha = \scp{\phi_\alpha}{\psi(0)}
\ee
then
\be
\psi(t) = \sum_{\alpha=1}^\D e^{-iE_\alpha t} c_\alpha\, \phi_\alpha\,.
\ee
Thus,
\begin{align}
\overline{\scp{\psi(t)}{P_{eq}|\psi(t)}} 
&= \sum_{\alpha,\beta=1}^\D \overline{e^{i(E_\alpha-E_\beta)t}} 
\, c^*_\alpha \, c_\beta
\scp{\phi_\alpha}{P_{eq}|\phi_\beta}\,.
\end{align}
If $H$ is non-degenerate (which is the generic case) then $E_\alpha-E_\beta$ vanishes only for $\alpha=\beta$, so the time averaged exponential is $\delta_{\alpha\beta}$, and
\be\label{evalave}
\overline{\scp{\psi(t)}{P_{eq}|\psi(t)}} 
= \sum_{\alpha=1}^\D \bigl|c_\alpha\bigr|^2 
\scp{\phi_\alpha}{P_{eq}|\phi_\alpha}\,.
\ee
Thus, for the system to be in thermal equilibrium most of the time it is necessary and sufficient that the right hand side of \eqref{evalave} is close to 1. 

Now if an energy eigenstate $\phi_\alpha$ is not itself in thermal equilibrium then when $\psi(0)=\phi_\alpha$ the system is never in thermal equilibrium, since this state is stationary. Conversely, if we have that
\be\label{good}
\scp{\phi_\alpha}{P_{eq}|\phi_\alpha}\approx 1 \quad \text{for all } \alpha\,,
\ee
then the system will be in thermal equilibrium most of the time for all $\psi(0)$. This follows directly from \eqref{evalave} since the right hand side of \eqref{evalave} is an average of the $\scp{\phi_\alpha}{P_{eq}|\phi_\alpha}$. We show below that \eqref{good} is true of ``most'' Hamiltonians, and thus, for ``most'' Hamiltonians it is the case that every wave function spends most of the time in thermal equilibrium.

\subsection{Main result}

The measure of ``most'' we use is the following: for any given $\D$ (distinct) energy values $E_1,\ldots,E_\D$, we consider the uniform distribution $\mu_\mathrm{Ham}$ over all Hamiltonians with these eigenvalues. Choosing $H$ at random with distribution $\mu_\mathrm{Ham}$ is equivalent to choosing the eigenbasis $\{\phi_\alpha\}$ according to the uniform distribution $\mu_{ONB}$ over all orthonormal bases of $\Hilbert$, and setting $H=\sum_\alpha E_\alpha |\phi_\alpha\rangle\langle\phi_\alpha|$. The measure $\mu_{ONB}$ can be defined as follows: Choosing a random basis according to $\mu_{ONB}$ amounts to choosing $\phi_1$ according to the uniform distribution over the unit sphere in $\Hilbert$, then $\phi_2$ according to the uniform distribution over the unit sphere in the orthogonal complement of $\phi_1$, etc. Alternatively, $\mu_{ONB}$ can be defined in terms of the Haar measure $\mu_{\Unitary(\D)}$ on the group $\Unitary(\D)$ of unitary $\D\times\D$ matrices: any given orthonormal basis $\{\chi_\alpha\}$ of $\Hilbert$ defines a one-to-one correspondence between $\Unitary(\D)$ and the set of all orthonormal bases of $\Hilbert$, associating with the matrix $U=(U_{\alpha\beta})\in\Unitary(\D)$ the basis
\be\label{phichi}
\phi_\alpha = \sum_{\beta=1}^\D U_{\alpha\beta} \chi_\beta\,;
\ee 
the image of the Haar measure under this correspondence is in fact independent of the choice of $\{\chi_\beta\}$ (because of the invariance of the Haar measure under right multiplication), and is $\mu_{ONB}$.

Put differently, the ensemble $\mu_\mathrm{Ham}$ of Hamiltonians can be obtained by starting from a given Hamiltonian $H_0$ on $\Hilbert$ (with distinct eigenvalues $E_1,\ldots,E_\D$) and setting
\be\label{HUHU}
H=UH_0U^{-1}
\ee
with $U$ a random unitary matrix chosen according to the Haar measure. Note that, while considering different possible Hamiltonians $H$ in $\Hilbert$, we keep $\Hilbert_{eq}$ fixed, although in practice it would often be reasonable to select $\Hilbert_{eq}$ in a way that depends on $H$ (as we did in the examples of Section~\ref{sec:example}).

For our purpose it is convenient to choose the basis $\{\chi_\alpha\}$ in such a way that the first $\dd_{eq}$ basis vectors lie in $\Hilbert_{eq}$ and the other ones are orthogonal to $\Hilbert_{eq}$. Then, we have that
\be\label{phialphaU}
\scp{\phi_\alpha}{P_{eq}|\phi_\alpha}=\sum_{\beta=1}^{\dd_{eq}} |U_{\alpha\beta}|^2
\ee
with $U_{\alpha\beta}$ the unitary matrix satisfying \eqref{phichi}.

We will show first, in Lemma 1, that for every $0<\varepsilon<1$, if $\D$ is sufficiently large and $\dd_{eq}/\D$ sufficiently close to 1, then most orthonormal bases $\{\phi_\alpha\}$ are such that 
\be\label{good2}
\scp{\phi_\alpha}{P_{eq}|\phi_\alpha}> 1-\varepsilon \quad \text{for all } \alpha\,.
\ee
This inequality is a precise version of \eqref{good}. How close to 1 should $\dd_{eq}/\D$ be? The fact that the average of $\scp{\psi}{P_{eq}|\psi}$ over all wave functions $\psi$ on the unit sphere of $\Hilbert$ equals $\dd_{eq}/\D$, mentioned already in \eqref{eq:mostVectors}, implies that \eqref{good2} cannot be true of most orthonormal bases if $\dd_{eq}/\D \leq 1-\varepsilon$. To have enough wiggling room, we require that
\be\label{deqepsilon}
\frac{\dd_{eq}}{\D} > 1- \frac{\varepsilon}{2}\,.
\ee

We will show then, in Theorem 1, that for every (arbitrarily small) $0<\eta<1$ and for sufficiently large $\D$, most $H$ are such that for every initial wave function $\psi(0)\in\Hilbert$ with $\|\psi(0)\|=1$, the system will spend most of the time in thermal equilibrium with accuracy $1-\eta$, where we say that a system with wave function $\psi$ is in thermal equilibrium with accuracy $1-\eta$ if
\be\label{eq2}
\scp{\psi}{P_{eq}|\psi}>1-\eta\,.
\ee
This inequality is a precise version of \eqref{eq1}. In order to have no more exceptions in time than the fraction $0<\delta'<1$, we need to set the $\varepsilon$ in \eqref{good2} and \eqref{deqepsilon} equal to $\eta\delta'$.

\begin{lem}\label{lem:typicalU}
Let $\mu_{\Unitary(\D)}$ denote the Haar measure on $\Unitary(\D)$, and
\be
\Set_{\varepsilon} := \Bigl\{U\in\Unitary(\D)\Big| \forall \alpha:\:
\sum_{\beta=1}^{\dd_{eq}} |U_{\alpha\beta}|^2>1-\varepsilon \Bigr\}\,.
\ee
Then for all $0<\varepsilon <1$ and $0<\delta<1$, there exists $\D_0=\D_0(\varepsilon,\delta)>0$ such that
\be
\text{if $\D>\D_0$ and $\dd_{eq}>(1-\varepsilon/2)\D$ then }\mu_{\Unitary(\D)}(\Set_{\varepsilon})\geq 1-\delta\,. 
\ee
\end{lem}

The proof of Lemma~\ref{lem:typicalU} is given in Section~\ref{sec:propStationary}. It also shows that $\D_0$ can for example be chosen to be
\be\label{D0}
\D_0(\varepsilon,\delta)=
\max\Bigl(10^3 \varepsilon^{-2} \log(4/\delta), 10^6 \varepsilon^{-4}\Bigr)\,.
\ee
From \eqref{phialphaU}, we obtain:

\begin{thm}\label{thm:1}
For all $\eta,\delta,\delta'\in(0,1)$, all integers $\D>\D_0(\eta\delta',\delta)$ and all integers $\dd_{eq}>(1-\eta\delta'/2)\D$ the following is true: Let $\Hilbert$ be a Hilbert space of dimension $\D$; let $\Hilbert_{eq}$ be a subspace of dimension $\dd_{eq}$; let $P_{eq}$ denote the projection to $\Hilbert_{eq}$; let $E_1,\ldots,E_\D$ be pairwise distinct but otherwise arbitrary; choose a Hamiltonian at random with eigenvalues $E_\alpha$ and an eigenbasis $\phi_\alpha$ that is uniformly distributed. Then, with probability at least $1-\delta$, every initial quantum state will spend $(1-\delta')$-most of the time in thermal equilibrium as defined in \eqref{eq2}, i.e.,
\be
\liminf_{T\to\infty} \frac{1}{T} \Bigl|\bigl\{0<t<T: \scp{\psi(t)}{P_{eq}|\psi(t)}>1-\eta\bigr\}\Bigr| \geq 1-\delta'\,,
\ee
where $|\set|$ denotes the size (Lebesgue measure) of the set $\set$.
\end{thm}

\proof
It follows from Lemma~\ref{lem:typicalU} that under the hypotheses of Theorem~\ref{thm:1}, $$\overline{\scp{\psi(t)}{P_{eq}|\psi(t)}} \geq 1- \eta\delta'$$ with probability at least $1-\delta$. Thus, since $\eta\delta'\geq\overline{1- \scp{\psi(t)}{P_{eq}|\psi(t)}}\geq\eta\tilde\delta$, where $\tilde{\delta}$ is the $\limsup_{T\to\infty}$ of the fraction of
the time in $(0,T)$ for which $\scp{\psi(t)}{P_{eq}|\psi(t)} \leq 1-\eta$,
it follows that $\ \tilde\delta\leq\delta'$.
\endproof

\subsection{Remarks}
\label{sec:remarks}

\textit{Normal typicality.}
Theorem~\ref{thm:1} can be strengthened; with the same sense of ``most'' as in Theorem~\ref{thm:1}, we have that for most Hamiltonians and for all $\psi(0)$
 \be\label{approx1}
\scp{\psi(t)}{P_{\nu}|\psi(t)} \approx \frac{\dim\Hilbert_\nu}{\dim \Hilbert}\,, \quad \text{for all }\nu
\ee
for most $t$. For $\nu=eq$, this implies that $\scp{\psi(t)}{P_{eq}|\psi(t)} \approx 1$. This stronger statement we have called normal typicality \cite{GLMTZ09}. A version of normal typicality was proven by von Neumann \cite{vN29}. However, because of the technical assumptions he made, von Neumann's result, while much more difficult, does not quite cover the simple result of this paper.

\bigskip

\textit{Typicality and probability.} When we express that something is true for most $H$ or most $\psi$ relative to some normalized measure $\mu$, it is often convenient to use the language of probability theory and speak of a random $H$ or $\psi$ chosen with distribution $\mu$. However, by this we do not mean to imply that the actual $H$ or $\psi$ in a concrete physical situation is random, nor that one would obtain, in repetitions of the experiment or in a class of similar experiments, different $H$'s or $\psi$'s whose empirical distribution is close to $\mu$. That would be a misinterpretation of the measure $\mu$, one that suggests the question whether perhaps the actual distribution in reality could be non-uniform. This question misses the point, as there need not be any actual distribution in reality. Rather, Theorem 1 means that the set of ``bad'' Hamiltonians has very small measure $\mu_\mathrm{Ham}$.

\bigskip

\textit{Consequences for Example 2.} From Lemma 1 it follows for Example 2 that typical Hamiltonians of the form \eqref{HUHU} with $H_0$ given by the right hand side of \eqref{Hex2} are such that all eigenfunctions are close to $\Hilbert_{0.5}$; this fact in turn strongly suggests (though we have not proved this) that the eigenfunctions are essentially concentrated on those configurations that have approximately $50\%$ of the particles in the left half and $50\%$ in the right half of the box.

\bigskip

\textit{Equilibrium Statistical Mechanics.}
Theorem 1 implies that, for typical $H$, every $\psi(0)\in\Hilbert$ is such that for most $t$, 
\be\label{psitMi}
\scp{\psi(t)}{M_i|\psi(t)} \approx \Tr(\rho_{mc} M_i)\,,
\ee
where $\rho_{mc}$ is the standard micro-canonical density matrix (i.e., $1/\D$ times the projection $\Hilbert_{\mathrm{total}}\to\Hilbert$), for all macro-observables $M_i$ as described in Section~\ref{sec:subspace}. This justifies replacing $\ket{\psi(t)}\bra{\psi(t)}$ by $\rho_{mc}$ as far as macro-observables in equilibrium are concerned. However, this does not, by itself, justify the use of $\rho_{mc}$ for observables $A$ not among the $\{M_i\}$. For example, consider a microscopic observable $A$ that is not ``nearly constant'' on the energy shell $\Hilbert$. Then, standard equilibrium statistical mechanics tells us to use $\rho_{mc}$ for the expected value of $A$ in equilibrium. We believe that this is in fact correct for most such observables, but it is not covered by Theorem 1. Results concerning many such observables are described in Section~\ref{sec:history}. These results, according to which, in an appropriate sense,
\be
\scp{\psi(t)}{A|\psi(t)} \approx \Tr(\rho_{mc}A)
\ee
for suitable $A$ and $\psi(0)$, are valid only in quantum mechanics. The justification of the broad use of $\rho_{mc}$ in classical statistical mechanics relies on rather different sorts of results requiring different kinds of considerations.

\section{Proof of Lemma~\ref{lem:typicalU}}
\label{sec:propStationary}

\label{sec:ProofLemma}

\proof 
Let us write $\PPP$ for the Haar measure $\mu_{\Unitary(\D)}$, and let
\be
p:= \PPP\Bigl(\bigcap_{\alpha=1}^\D \Bigl\{
\sum_{\beta=1}^{d_{eq}}|U_{\alpha\beta}|^2 >1-\varepsilon\Bigr\}\Bigr)\,.
\ee
Observe that
\begin{align}
p &= 1 - \PPP\Bigl(\bigcup_{\alpha=1}^\D \Bigl\{\sum_{\beta=1}^{d_{eq}}|U_{\alpha\beta}|^2
\leq 1- \varepsilon\Bigr\}\Bigr)\\
&\geq 1-\D\,\max_{\alpha} \PPP\Bigl\{\sum_{\beta=1}^{\dd_{eq}}
|U_{\alpha\beta}|^2\leq 1- \varepsilon\Bigr\}\,.
\end{align}
Since $U=(U_{\alpha\beta})$ is a random unitary matrix with Haar distribution, its $\alpha$-th column is a random unit vector $\vec{U}:=(U_{\alpha\beta})_\beta$ whose distribution is uniform over the unit sphere of $\CCC^\D$ (i.e., the distribution is, up to a normalizing constant, the surface area measure). Therefore, the probability in the last line does not, in fact, depend on $\alpha$, and so the step of taking the maximum over $\alpha$ can be omitted.

A random unit vector such as $\vec{U}$ can be thought of as arising from a random Gaussian vector $\vec{G}$ by normalization: Let $G_\beta$ for $\beta=1,\ldots,\D$ be independent complex Gaussian random variables with mean 0 and variance $\EEE|G_\beta|^2=1/\D$; i.e., $\Re\,G_\beta$ and $\Im\,G_\beta$ are independent real Gaussian random variables with mean 0 and variance $1/2\D$. Then the distribution of $\vec{G}=(G_1,\ldots,G_\D)$ is symmetric under rotations from $\Unitary(\D)$, and thus
\be
\frac{\vec{G}}{\|\vec{G}\|} = \vec{U} \text{ in distribution, with }
\|\vec{G}\|^2=\sum_{\beta=1}^\D |G_\beta|^2\,.
\ee
We thus have that
\be\label{eqInequality2}
p\geq 1- \D\, \PPP\Bigl\{\sum_{\beta=1}^{\dd_{eq}}
\frac{|G_{\beta}|^2}{\|\vec{G}\|^2}\leq 1- \varepsilon\Bigr\}\,.
\ee

To estimate the probability on the right hand side of
(\ref{eqInequality2}), we introduce three different events:
\begin{align}
A(\eta') &:= \Bigl\{\bigl|\|\vec{G}\|^2 - 1\bigr| < \eta' \Bigr\}\, ,\\
B(\eta'') &:= \Bigl\{ (1-\eta'')\frac{\dd_{eq}}{\D}
<\sum_{\beta=1}^{\dd_{eq}} |G_\beta|^2
<(1+\eta'')\frac{\dd_{eq}}{\D} \Bigr\}\, ,\\
C(\eta''') &:=\Bigl\{ (1-\eta''')\frac{\dd_{eq}}{\D}< \sum_{\beta=1}^{\dd_{eq}}
\frac{|G_\beta|^2}{\|\vec{G}\|^2} <(1+\eta''') \frac{\dd_{eq}}{\D} \Bigr\}\, .
\end{align}
Let us now assume that
\be
\frac{\dd_{eq}}{\D}> 1-\frac{\varepsilon}{2}\,.
\ee
We then have that
\be
(1-\varepsilon/2)\frac{\dd_{eq}}{\D}>1-\varepsilon+\frac{\varepsilon^2}{4}>1-\varepsilon\,,
\ee
so that
\be
C(\varepsilon/2) 
\subseteq \Bigl\{(1-\varepsilon/2)\frac{\dd_{eq}}{\D} < \sum_{\beta=1}^{\dd_{eq}}
\frac{|G_\beta|^2}{\|\vec{G}\|^2}\Bigr\}
\subseteq \Bigl\{1-\varepsilon < \sum_{\beta=1}^{\dd_{eq}}
\frac{|G_\beta|^2}{\|\vec{G}\|^2}\Bigr\}
\ee
and thus
\be\label{eqEquality1}
p\geq 1-\D\,\PPP(C^c(\varepsilon/2))\,,
\ee
where the superscript $c$ means complement. Our goal is to find a good upper bound for $\PPP(C^c(\varepsilon/2))$.

If the event $A(\eta')$ occurs for $0<\eta'<\frac12$ then
\begin{equation}
1-\eta' < \frac{1}{\|\vec{G}\|^2} < 1+2\eta'\,, 
\end{equation}
and consequently, if $A(\eta')\cap B(\eta'')$ occurs then
\begin{equation}
\frac{\dd_{eq}}{\D}(1-\eta')(1-\eta'') < \frac{\sum\limits_{\beta=1}^{\dd_{eq}}
|G_\beta|^2}{\|\vec{G}\|^2} <
\frac{\dd_{eq}}{\D}(1+2\eta')(1+\eta'')\, .
\end{equation}
It is now easy to see that $A(\eta')\cap B(\eta'') \subseteq
C(2\eta'+\eta''+2\eta'\eta'')$, so if we choose $\eta' = \eta'' =
\varepsilon/8$ we obtain that
\begin{equation}
A(\tfrac{\varepsilon}{8})\cap B(\tfrac{\varepsilon}{8}) \subseteq 
C(\tfrac{3}{8}\varepsilon + \tfrac{1}{32}\varepsilon^2)\subseteq C(\varepsilon/2) 
\ \ \ \text{for} \ \ \ 0<\varepsilon < 1\,.
\end{equation}
We thus have the following upper bound:
\begin{equation}\label{eqInequality3}
\PPP(C^c(\varepsilon/2)) \leq \PPP(A^c(\varepsilon/8)) + \PPP(B^c(\varepsilon/8))\,.
\end{equation}

\bigskip

To find an estimate of $\PPP(A(\varepsilon/8))$ and $\PPP(B(\varepsilon/8))$ we use the \textit{Large Deviations Principle}. It is convenient to use a slightly stronger version of this principle than usual, see Section 2.2.1 of \cite{DZ98}, which states that for a sequence of $N$ i.i.d.\ random variables $X_i$,
\begin{equation}\label{largeDev}
\PPP\Bigl(\bigl|\sum_{i=1}^N \frac{X_i}{N} - \EEE(X_1)\bigr| >
\delta\Bigr) \leq 2e^{-NI(\EEE(X_1) + \delta)}
\end{equation}
where $I(x)$ is the \emph{rate function} \cite{DZ98} associated with the distribution of the $X_i$, defined to be
\be
I(x)=\sup_{\theta>0} (\theta x-\log \EEE e^{\theta X_i})\,.
\ee
In our case, where $X_i$ will be the square of a standard normal random variable, the rate function is
\be\label{rateFunc}
I(x) = \tfrac{1}{2}(x-1-\log x) \quad \forall x>1\,,
\ee
as a simple calculation shows.

To estimate $\PPP(A(\varepsilon/8))$, set
\be
N=2\D\,,\quad
X_\beta = 2\D (\Re\, G_\beta)^2\,,\quad
X_{\D+\beta} = 2\D (\Im\, G_\beta)^2 \,\quad
\text{for }\beta=1,\ldots,\D\,. 
\ee
Thus, for $i=1,\ldots,2\D$, the $X_i$ are i.i.d.\  variables with mean $\EEE X_i=2\D\, \EEE(\Re\, G_i)^2=1$; we thus obtain
\begin{align}
\PPP(A^c(\varepsilon/8)) &= \PPP\Bigl\{
\bigl|\|\vec{G}\|^2 - 1\bigr| >\varepsilon/8 \Bigr\} = \\
&= \PPP\Bigl\{\bigl| \sum_{\beta=1}^\D |G_\beta|^2 -
1\bigr|>\varepsilon/8 \Bigr\} \\
&= \PPP\Bigl\{\bigl| \sum_{i=1}^{2\D} \frac{X_i}{2\D} -1\bigr|>\varepsilon/8 \Bigr\} \\
&\le 2 e^{-2\D\,I(1+\varepsilon/8)}\\
&= 2e^{-\D(\varepsilon/8-\log(1+\varepsilon/8))}\\
&\le 2 \exp\Bigl( - \frac{\D\varepsilon^2}{192}\Bigr)\,.\label{eqNormGauss2}
\end{align}
In the last step we have used that $\log(1+x)\leq x-x^2/3$ for $0<x<1/2$.

We use a completely analogous argument for $B$, setting
\be
N=2\dd_{eq}\,,\quad
X_\beta = 2\D (\Re\,G_\beta)^2\,,\quad
X_{\D+\beta} = 2\D (\Im\,G_\beta)^2\,,\quad
\text{for }\beta=1,\ldots,\dd_{eq}\,,
\ee
and obtain that
\begin{align}\label{eqNormGauss3}
\PPP(B^c(\varepsilon/8)) &= \PPP\Bigl\{
\Bigl|\sum_{\beta=1}^{\dd_{eq}} |G_\beta|^2 -
\frac{\dd_{eq}}{\D}\Bigr|/\frac{\dd_{eq}}{\D} > \varepsilon/8 \Bigr\}\\
&= \PPP\Bigl\{\bigl| \sum_{i=1}^{2\dd_{eq}} \frac{X_i}{2\dd_{eq}} -1\bigr|>\varepsilon/8 \Bigr\} \\
\label{eqNormGauss4}
&\leq  2 \exp\Bigl( -\frac{\dd_{eq}\varepsilon^2}{192}\Bigr)\,.
\end{align}

From (\ref{eqInequality3}), (\ref{eqNormGauss2}), and (\ref{eqNormGauss4}) it follows that
\begin{equation}
\PPP(C^c(\varepsilon/2)) \leq 2 \exp\Bigl( -
\frac{\dd_{eq}\varepsilon^2}{192}\Bigr) + 2 \exp\Bigl( -
\frac{\D\varepsilon^2}{192}\Bigr)\leq 4 \exp\Bigl( -
\frac{\D\varepsilon^2}{384}\Bigr)\, ,
\end{equation}
where we have used that $\dd_{eq}>\D/2$. Therefore, by \eqref{eqEquality1},
\be\label{estimate} 
p\geq 1-4\D\exp\Bigl(-\frac{\D\varepsilon^2}{384} \Bigr)\,.
\ee
The last term converges to 0 as $\D\to\infty$, so there exists a $\D_0>0$ such that for all $\D>\D_0$,
\be
p \geq 1-\delta\,,
\ee
which is what we wanted to show. In order to check this for the $\D_0$ specified in \eqref{D0} right after Lemma~\ref{lem:typicalU}, note that the desired relation
\be
4\D\exp\Bigl(-\frac{\D\varepsilon^2}{384} \Bigr) \leq \delta
\ee
is equivalent to
\be
\D\Bigl(\frac{\varepsilon^2}{384} -\frac{\log \D}{\D}\Bigr) \geq \log(4/\delta)\,.
\ee
Thus, it suffices that $\D>10^3 \varepsilon^{-2} \log(4/\delta)$ and 
\be\label{sufficient2}
\frac{\log \D}{\D} < 10^{-3}\varepsilon^{2}\,.
\ee
Since $\log \D<\sqrt{\D}$ for all positive numbers $\D$, condition \eqref{sufficient2} will be satisfied if $\sqrt{\D}>10^3 \varepsilon^{-2}$, i.e., if $D>10^6 \varepsilon^{-4}$. 
\endproof

\section{Examples of systems that do not approach thermal equilibrium}
\label{sec:badCouplings}

We shall now present examples of atypical behavior, namely examples of ``bad'' Hamiltonians, i.e., Hamiltonians for which not all wave functions approach thermal equilibrium (or, equivalently, for which \eqref{good} is not satisfied). According to Theorem~\ref{thm:1}, bad Hamiltonians form a very small subset of the set of all Hamiltonians. Of course, to establish that \eqref{good} holds for a particular Hamiltonian can be a formidable challenge. Moreover, the small subset might include all standard many-body Hamiltonians (e.g., all those which are a sum of kinetic and potential energies). But there is no a priori reason to believe that this should be the case. 

The first example consists of two non-interacting subsystems. This can be expressed in the framework provided by Example 1 in Section~\ref{sec:example} with the Hamiltonian $H=H_1+H_2+\lambda V$ by setting $\lambda=0$. Let $\{\phi_i^1\}$ be an orthonormal basis of $\Hilbert_1$ consisting of eigenvectors of $H_1$ with eigenvalues $E_i^1$, and $\{\phi_j^2\}$ one of $\Hilbert_2$ consisting of eigenvectors of $H_2$ with eigenvalues $E_j^2$. Clearly, for $\lambda=0$ not every wave function will approach thermal equilibrium. After all, in this case, the $\phi_i^1\otimes \phi_j^2$ form an eigenbasis of $H$, while
\be\label{Hilbertlambda0}
\Hilbert = \text{span} \bigl\{ \phi_i^1\otimes \phi_j^2: E_i^1+E_j^2\in [E,E+\delta E]\bigr\}
\ee
and
\be\label{Hilberteqlambda0}
\Hilbert_{eq} =\text{span} \bigl\{\phi_i^1\otimes \phi_j^2: E_i^1 \in [0.49 E, 0.51 E) \text{ and } E_i^1+E_j^2 \in [E,E+\delta E] \bigr\}\,.
\ee
Thus, any $\phi_i^1\otimes \phi_j^2$ such that $E_i^1+E_j^2 \in [E,E+\delta E]$ but, say, $E_i^1< 0.49 E$, will be an example of an element of $\Hilbert$ that is orthogonal to $\Hilbert_{eq}$ and, as it is an eigenfunction of $H$, forever remains orthogonal to $\Hilbert_{eq}$. 

As another example, we conjecture that some wave functions will fail to approach thermal equilibrium also when $\lambda$ is nonzero but sufficiently small. We prove this now for a slightly simplified setting, corresponding to the following modification of Example 1 of Section~\ref{sec:example}. For the usual energy interval $[E,E+\delta E]$, let $\Hilbert$ be, independently of $\lambda$, given by \eqref{Hilbertlambda0}, and, instead of $H_1+H_2+\lambda V$, let $H$ be given by
\be
H=H(\lambda)=P(H_1+H_2+\lambda V)P\,,
\ee
where $P$ is the projection to $\Hilbert$. Then $H$ defines a time evolution on $\Hilbert$ that depends on $\lambda$. (Note that $\Hilbert$ is still an ``energy shell'' for all sufficiently small $\lambda$, as all nonzero eigenvalues of $H(\lambda)$ are still contained in an interval just slightly larger than $[E,E+\delta E]$, and the corresponding eigenvectors lie in $\Hilbert$.) Let $\Hilbert_{eq}$ for $\lambda\neq 0$ also be given by \eqref{Hilberteqlambda0}. Again, choose one particular $\phi_i^1$ and one particular $\phi_j^2$ (independently of $\lambda$) so that $E_i^1+E_j^2\in [E,E+\delta E]$ and $E_i^1<0.49 E$, and consider as the initial state of the system again
\be
\psi(t=0)=\phi_i^1\otimes \phi_j^2\,,
\ee
which evolves to 
\be
\psi(\lambda,t) = e^{-iH(\lambda)t} \phi_i^1\otimes\phi_j^2\,.
\ee
Suppose for simplicity that $H(\lambda=0)=H_1+H_2$ is non-degenerate.\footnote{Since this requires that no eigenvalue difference of $H_1$, $E^1_i-E^1_{i'}$, coincides with an eigenvalue difference of $H_2$, $E^2_j-E^2_{j'}$, we need to relax our earlier assumption that system 1 and system 2 be identical; so, let them be almost identical, with slightly different eigenvalues, and let $H_1$ and $H_2$ each be non-degenerate.} Then, according to standard results of perturbation theory \cite{Kato}, also $H(\lambda)$, regarded as an operator on $\Hilbert$, is non-degenerate for all $\lambda\in(-\lambda_0,\lambda_0)$ for some $\lambda_0>0$; moreover, its eigenvalues $E(\lambda)$ depend continuously (even analytically) on $\lambda$, and so do the eigenspaces. In particular, it is possible to choose for every $\lambda\in(-\lambda_0,\lambda_0)$ a normalized eigenstate $\phi(\lambda)\in\Hilbert$ of $H(\lambda)$ with eigenvalue $E(\lambda)$ in such a way that $\phi(\lambda)$ and $E(\lambda)$ depend continuously on $\lambda$, and $\phi(\lambda=0)=\phi_i^1\otimes \phi_j^2$.

We are now ready to show that for sufficiently small $\lambda>0$,
\be\label{nearortho}
\scp{\psi(\lambda,t)}{P_{eq}|\psi(\lambda,t)} \approx 0
\ee
for all $t$; that is, $\psi(\lambda,t)$ is nearly orthogonal to $\Hilbert_{eq}$ for all $t$, and thus is never in thermal equilibrium. To see this, note first that since $\phi(0) \approx \phi(\lambda)$ for sufficiently small $\lambda$, and since $e^{-iH(\lambda)t}$ is unitary, also
\be\label{psitphilambda}
e^{-iH(\lambda)t} \phi(0) \approx e^{-iH(\lambda)t} \phi(\lambda)
\ee
(with error independent of $t$). Since the right hand side equals
\be
e^{-iE(\lambda)t}\phi(\lambda)\approx e^{-iE(\lambda)t} \phi(0)\,,
\ee
we have that
\be\label{approx2}
\scp{e^{-iH(\lambda)t} \phi(0)}{P_{eq}|e^{-iH(\lambda)t} \phi(0)} \approx 
\scp{\phi(0)}{P_{eq}|\phi(0)}=0\,.
\ee
This proves \eqref{nearortho} with an error bound independent of $t$ that tends to 0 as $\lambda\to 0$.

Another example of ``bad'' Hamiltonians is provided by the phenomenon of \emph{Anderson localization} (see in particular \cite{And58,OH07}): Certain physically relevant Hamiltonians possess some eigenfunctions $\phi_\alpha$ that have a spatial energy density function that is macroscopically non-uniform whereas wave functions in $\Hilbert_{eq}$ should have macroscopically uniform energy density over the entire available volume. Thus, some eigenfunctions are not close to $\Hilbert_{eq}$, violating \eqref{good}.

\section{Comparison with classical mechanics}

In classical mechanics, one would expect as well that a macroscopic system spends most of the time in the long run in thermal equilibrium. Let us define what thermal equilibrium means in classical mechanics. (We defined it for quantum systems in \eqref{eq1}.) We denote a point in phase space by $X=(\vq_1,\ldots,\vq_N,\vp_1,\ldots,\vp_N)$. Instead of the orthogonal decomposition of $\Hilbert$ into subspaces $\Hilbert_\nu$ we consider a partition of an energy shell $\Gamma$ in phase space, $\Gamma = \{X:E\leq H(X)\leq E+\delta E\}$, into regions $\Gamma_\nu$ corresponding to different macro-states $\nu$, i.e., if the micro-state $X$ of the system is in $\Gamma_\nu$ then the macro-state of the system is $\nu$. It has been shown \cite{Lan} for realistic systems with large $N$ that one of the regions $\Gamma_\nu$, corresponding to the macro-state of thermal equilibrium and denoted $\Gamma_{eq}$, is such that, in terms of the (uniform or Liouville) phase space volume measure $\mu$ on $\Gamma$,
\be
\frac{\mu(\Gamma_{eq})}{\mu(\Gamma)} \approx 1\,. 
\ee
Though the subspaces $\Hilbert_\nu$ play a role roughly analogous to the regions $\Gamma_\nu$, a basic difference between the classical and the quantum cases is that while every classical phase point in $\Gamma$ belongs to one and only one $\Gamma_\nu$, and thus is in one macro-state, a quantum state $\psi$ need not lie in any one $\Hilbert_\nu$, but can be a non-trivial superposition of vectors in different macro-states. (Indeed, almost all $\psi$ do not lie in any one $\Hilbert_\nu$. That is why we defined being in thermal equilibrium in terms of $\psi$ lying in a neighborhood of $\Hilbert_{eq}$, rather than lying in $\Hilbert_{eq}$ itself.)

The time evolution of the micro-state $X$ is given by the solution of the Hamiltonian equations of motion, which sends $X$ (at time $0$) to $X_t$ (at time $t$), $t\in\RRR$. We expect that for realistic systems with a sufficiently large number $N$ of constituents and for every macro-state $\nu$, most initial phase points $X\in\Gamma_\nu$ will be such that $X_t$ spends most of the time in the set $\Gamma_{eq}$. This statement follows if the system is ergodic,\footnote{A classical system is ergodic if and only if the time evolved micro-state $X_t$ spends, in the long run, a fraction of time in each (measurable) set $B\subseteq \Gamma$ that is equal to $\mu(B)/\mu(\Gamma)$ for $\mu$-almost all $X$.} but in fact is much weaker than ergodicity. Theorem~\ref{thm:1} is parallel to this statement in that it implies, for typical Hamiltonians, that initial states (here, $\psi(0)$) out of thermal equilibrium will spend most of the time in thermal equilibrium; it is different in that it applies, for typical Hamiltonians, to \emph{all}, rather than most, initial states $\psi(0)$.

\section{Comparison with the literature}
\label{sec:history}

Von Neumann \cite{vN29} proved, as his ``quantum ergodic theorem,'' a precise version of normal typicality (defined in Section~\ref{sec:remarks}); his proof requires much more effort, and more refined methods, than our proof of Theorem 1. However, his theorem assumes that the dimension $\dd_\nu$ of each macro-space $\Hilbert_\nu$ is much smaller than the full dimension $\D$, and thus does not apply to the situation considered in this paper, in which one of the macro-spaces, $\Hilbert_{eq}$, has the majority of dimensions. The reason von Neumann treated the more difficult case of small $\dd_\nu$ but left out the easier and particularly interesting case of the thermal equilibrium macrostate is that he had in mind a notion of thermal equilibrium different from ours. He thought of a thermal equilibrium wave function $\psi$, not as one in (or close to) a particular $\Hilbert_\nu$, but as one with $\|P_\nu\psi\|^2\approx \dd_\nu/\D$ for every $\nu$, i.e., one for which $|\psi\rangle\langle\psi|\approx \rho_{mc}$ in a suitable coarse-grained sense. Because of this different focus, he did not consider the situation presented here. We also note that von Neumann's quantum ergodic theorem makes an assumption on $H$ that we do not need in our Theorem 1; this assumption, known as a ``no resonances'' \cite{J63,T98} or ``non-degenerate energy gaps'' \cite{LPSW08} condition, asserts that 
\be\label{energygaps}
E_\alpha-E_\beta \neq E_{\alpha'} - E_{\beta'}\text{ unless }
\begin{cases}\text{either }\alpha=\alpha', \beta=\beta'\\
\text{or }\alpha=\beta, \alpha'=\beta'\,.\end{cases}
\ee

\bigskip

The Schnirelman Theorem \cite{C85} states that, in the semi-classical limit and under suitable hypotheses, the Wigner distribution corresponding to an eigenstate $\phi_\alpha$ becomes the micro-canonical measure. That is, the $\phi_\alpha$ have a property resembling thermal equilibrium, similar to our condition \eqref{good} expressing that all eigenstates are in thermal equilibrium. Srednicki \cite{Sre94} observed other thermal equilibrium properties in energy eigenstates of example systems, a phenomenon he referred to as ``eigenstate thermalization.''

\bigskip

The results of \cite{T98,R08,LPSW08} also concern conditions under which a quantum system will spend most of the time in ``thermal equilibrium.'' For the sake of comparison, their results, as well as ours, can be described in a unified way as follows. Let us say that a system with initial wave function $\psi(0)$ \emph{equilibrates} relative to a class $\A$ of observables if for most times $\tau$,
\be\label{equidef}
\scp{\psi(\tau)}{A|\psi(\tau)} \approx 
\Tr\Bigl(\overline{\ket{\psi(t)}\bra{\psi(t)}}A\Bigr) 
\text{ for all }A\in\A\,.
\ee
We then say that the system \emph{thermalizes} relative to $\A$ if it equilibrates and, moreover,
\be
\Tr\Bigl(\overline{\ket{\psi(t)}\bra{\psi(t)}} A\Bigr)\approx
\Tr\bigl(\rho_{mc}A\bigr) \text{ for all }A\in\A\,,
\ee
with $\rho_{mc}$ the micro-canonical density matrix (in our notation, $1/\D$ times the projection $P$ to $\Hilbert$). With these definitions, the results of \cite{T98,R08,LPSW08} can be formulated by saying that, under suitable hypotheses on $H$ and $\psi(0)$ and for large enough $\D$, a system will equilibrate, or even thermalize, relative to a suitable class $\A$.

Our result is also of this form. We have just one operator in $\A$, namely $P_{eq}$. We establish thermalization for arbitrary $\psi(0)$ assuming $H$ is non-degenerate and satisfies $\scp{\phi_\alpha}{P_{eq}|\phi_\alpha}\approx 1$ for all $\alpha$, which (we show) is typically true.

Von Neumann's quantum ergodic theorem \cite{vN29} establishes thermalization for a family $\A$ of commuting observables; $\A$ is the algebra generated by $\{M_1,\ldots,M_k\}$ in the notation of Section~\ref{sec:subspace}. He assumes that the dimensions of the joint eigenspaces $\Hilbert_\nu$ are not too small and not too large; that $H$ obeys \eqref{energygaps}; he makes an assumption about the relation between $H$ and the subspaces $\Hilbert_\nu$ that he shows is typically true; and he admits arbitrary $\psi(0)$. See \cite{GLMTZ09} for further discussion. Rigol, Dunjko, and Olshanii \cite{RDO08} numerically simulated an example system and concluded that it thermalizes relative to a certain class $\A$ consisting of commuting observables.

Tasaki \cite{T98} as well as Linden, Popescu, Short, and Winter \cite{LPSW08} consider a system coupled to a heat bath, $\Hilbert_\mathrm{total}=\Hilbert_\mathrm{sys}\otimes\Hilbert_\mathrm{bath}$, and take $\A$ to contain all operators of the form $A_\mathrm{sys}\otimes 1_\mathrm{bath}$. Tasaki considers a rather special class of Hamiltonians and establishes thermalization assuming that 
\be
\max_\alpha |\scp{\phi_\alpha}{\psi(0)}|^2 \ll 1\,,
\ee
a condition that implies that many eigenstates of $H$ contribute to $\psi(0)$ appreciably and that can (more or less) equivalently be rewritten as
\be\label{contribute}
\sum_\alpha \bigl|\scp{\phi_\alpha}{\psi(0)}\bigr|^4 \ll 1\,.
\ee
Under the assumption \eqref{contribute} on $\psi(0)$,
Linden et al.\ establish equilibration for $H$ satisfying \eqref{energygaps}. They also establish a result in the direction of thermalization under the additional hypothesis that the dimension of the energy shell of the bath is much greater than $\dim \Hilbert_\mathrm{sys}$.

Reimann's mathematical result \cite{R08} can be described in the above scheme as follows. Let $\A$ be the set of all observables $A$ with (possibly degenerate) eigenvalues between 0 and 1 such that the absolute difference between any two eigenvalues is at least (say) $10^{-1000}$. He establishes equilibration for $H$ satisfying \eqref{energygaps}, assuming that $\psi(0)$ satisfies \eqref{contribute}.

\bigskip

\noindent\textit{Acknowledgements.}
We thank Matthias Birkner (LMU M\"unchen), Peter Reimann (Bielefeld), Anthony Short (Cambridge), Avraham Soffer (Rutgers), and Eugene Speer (Rutgers) for helpful discussions. 
S.~Goldstein was supported in part by National Science Foundation [grant DMS-0504504].
N.~Zangh\`\i\ is supported in part by Istituto Nazionale di Fisica Nucleare. 
J.~L.~Lebowitz and C.~Mastrodonato are supported in part by NSF [grant DMR 08-02120] and by AFOSR [grant AF-FA 09550-07].


\begin{thebibliography}{14}

\bibitem{And58} P. W. Anderson:
	Absence of Diffusion in Certain Random Lattices.
	\textit{Phys. Rev.} \textbf{109}, 1492--1505, 1958.
	
\bibitem{C85} Y. Colin de Verdi\`ere:
                            Ergodicit\'e et fonctions propres du
                            Laplacien.  \textit{Commun. Math. Phys.} \textbf{102},
                            497--502, 1985.

\bibitem{DZ98} A. Dembo, O. Zeitouni:
	\textit{Large deviations techniques and applications}, 2nd ed.
	Springer, New York, 1998.

\bibitem{Deu91} J. M. Deutsch:
	Quantum statistical mechanics in a closed system.
	\textit{Phys. Rev. A} \textbf{43}, 2046--2049, 1991.

\bibitem{GLMTZ09} 
S. Goldstein, J. L. Lebowitz, C. Mastrodonato, R. Tumulka, N. Zangh\`\i:
	Normal Typicality and von Neumann's Quantum Ergodic Theorem.
	\url{http://arxiv.org/abs/0907.0108}, 2009.

\bibitem{J63} R. Jancel: 
	\textit{Foundations of Classical and Quantum Statistical Mechanics}.
	Oxford: Pergamon, 1969.
	Translation by W. E. Jones of
	\textit{Les Fondements de la M\'ecanique Statistique Classique e Quantique.}
	Paris: Gauthier-Villars, 1963.

\bibitem{Kato} T. Kato:
	\textit{A short introduction to perturbation theory for linear operators.}
	New York: Springer-Verlag, 1982. 

\bibitem{Lan} O. E. Lanford: 
	Entropy and Equilibrium States in Classical Statistical Mechanics. 
	In A. Lenard (ed.), \textit{Lecture Notes in Physics} \textbf{2}, 1--113, 
	Springer-Verlag, 1973.

\bibitem{L07} J. L. Lebowitz:
	From Time-symmetric Microscopic Dynamics to Time-asymmetric 
	Macroscopic Behavior: An Overview.
	In G.~Gallavotti , W.~L.~Reiter, J.~Yngvason (editors), 
	\textit{Boltzmann's Legacy}, 63--88.
	European Mathematical Society (2007).
	\url{http://arxiv.org/abs/0709.0724}

\bibitem{LPSW08}
	N. Linden, S. Popescu, A. J. Short, A. Winter:
	Quantum mechanical evolution towards thermal equilibrium. 
	\textit{Phys. Rev E} \textbf{79}, 061103, 2009.
	\url{http://arxiv.org/abs/0812.2385}

\bibitem{OH07} V. Oganesyan, D. A. Huse:
	Localization of interacting fermions at high temperature.
	\textit{Phys. Rev. B} \textbf{75}, 155111, 2007. 

\bibitem{R08} P. Reimann: 
	Foundation of Statistical Mechanics under Experimentally Realistic Conditions. 
	\textit{Phys. Rev. Lett.} \textbf{101}, 190403, 2008.

\bibitem{RDO08} M. Rigol, V. Dunjko, M. Olshanii:
	Thermalization and its mechanism for generic isolated quantum systems.
	\textit{Nature} \textbf{452}, 854--858, 2008.

\bibitem{schrbook} E. Schr\"odinger: \textit{Statistical
  Thermodynamics.} Second Edition, Cambridge University Press, 1952.

\bibitem{Sre94} M. Srednicki:
	Chaos and quantum thermalization.
	\textit{Phys. Rev. E} \textbf{50}, 888, 1994.

\bibitem{T98} H. Tasaki: 
	From Quantum Dynamics to the Canonical
	Distribution: General Picture and a Rigorous Example. 
	\textit{Phys. Rev. Lett.} \textbf{80}, 1373-1376, 1998.

\bibitem{vN29} J. von Neumann:
      Beweis des Ergodensatzes und des $H$-Theorems in der neuen Mechanik. 
      \textit{Z. Physik} \textbf{57}, 30, 1929.

\bibitem{vN32} J. von Neumann:
                            \emph{Mathematical Foundation of Quantum Mechanics.} 
                            Princeton University Press, 1955.
                            Translation of \emph{Mathematische Grundlagen der
                            Quantenmechanik.} Springer-Verlag,
                            Berlin, 1932.

\bibitem{Bloch} J. D. Walecka: \textit{Fundamentals of
  Statistical Mechanics. Manuscript and Notes of Felix Bloch.}
  Stanford University Press, Stanford, CA, 1989.

\end{thebibliography}
\end{document}